\documentclass{article}
\usepackage{ijcai19}
\usepackage{times}
\usepackage{soul}
\usepackage{url}
\usepackage[utf8]{inputenc}
\usepackage[small]{caption}
\usepackage{graphicx}
\usepackage{amsmath}
\usepackage{booktabs}
\usepackage{subfig}
\usepackage{ijcai19}
\usepackage{times}
\usepackage{epsfig}
\usepackage{graphicx}
\usepackage{float}
\usepackage{amsmath}
\usepackage{amssymb}
\usepackage{subfig}
\usepackage{booktabs}
\usepackage{multirow}
\usepackage{enumerate}
\usepackage{array}
\usepackage{bm}
\renewcommand{\paragraph}{\textbf}

\title{Dynamically Visual Disambiguation of Keyword-based Image Search}

\author{
Yazhou Yao$^{\dag}$\and
Zeren Sun$^{\S}$\and
Fumin Shen$^{\natural}$\footnote{Corresponding Author}
Li Liu$^{\dag}$\and
Limin Wang$^{\sharp}$\and
Fan Zhu$^{\dag}$\and
Lizhong Ding$^{\dag}$\and
Gangshan Wu$^{\sharp}$\and
Ling Shao$^{\dag}$\\
\affiliations
$^\dag$Inception Institute of Artificial Intelligence, Abu Dhabi, UAE\\
$^\S$Nanjing University of Science and Technology, Nanjing, China\\
$^\natural$University of Electronic Science and Technology of China, Chengdu, China\\
$^\sharp$Nanjing University, Nanjing, China\\
\emails
\{yaoyazhou, zerensun, fumin.shen, liuli1213, lmwang.nju, fanzhu1987, lizhongding\}@gmail.com,\\
gswu@nju.edu.cn, ling.shao@ieee.org \\
}
\begin{document}

\maketitle

\begin{abstract}
Due to the high cost of manual annotation, learning directly from the web has attracted broad attention. One issue that limits their performance is the problem of visual polysemy. To address this issue, we present an adaptive multi-model framework that resolves polysemy by visual disambiguation. Compared to existing methods, the primary advantage of our approach lies in that our approach can adapt to the dynamic changes in the search results. Our proposed framework consists of two major steps: we first discover and dynamically select the text queries according to the image search results, then we employ the proposed saliency-guided deep multi-instance learning network to remove outliers and learn classification models for visual disambiguation. Extensive experiments demonstrate the superiority of our proposed approach.
\end{abstract}

\section{Introduction}

In the past few years, labeled image datasets have played a critical role in high-level image understanding \cite{simonyan2014very,min2016being,fang2018,xie2017sde,cvpr19aren,shu2018,wang2015instre,hu2017frankenstein,hu2017attribute,mta2018,huangpu,xu2017}. However, the process of constructing manually labeled datasets is both time-consuming and labor-intensive \cite{deng2009imagenet}. To reduce the time and labor cost of manual annotation, learning directly from the web images has attracted more and more attention \cite{chen2013neil,yao2018aaai,shen2018tmm,zhang2016domain,yaotkde2019,liutip2019,tangijcai2018,huatmm2017,xuicme2016,shu2015,prl2018,tang2017,mmm}.
Compared to manually-labeled image datasets, web images are a rich and free resource. For arbitrary categories, potential training data can be easily obtained from an image search engine \cite{zhang2016domain}. Unfortunately, the precision of returned images from an image search engine is still unsatisfactory. 

\begin{figure}[tbp]
	\centering
	\includegraphics[width=0.46\textwidth]{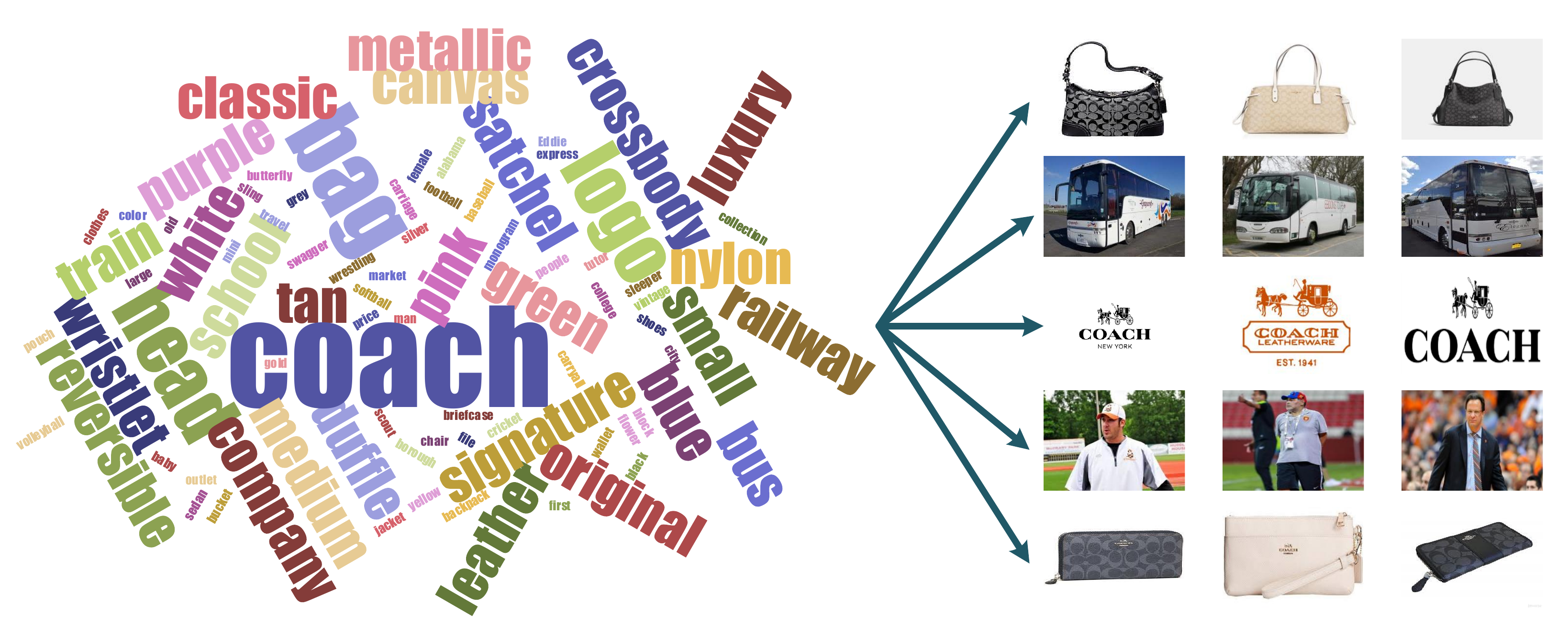}
	\caption{Visual polysemy. The keyword ``coach" can refer to multiple text semantics, resulting in images with various visual senses in the image search results.}
	\label{fig1} 
\end{figure}

One of the most important reasons for the noisy results is the problem of visual polysemy. As shown in Fig. \ref{fig1}, visual polysemy means that a word has multiple semantic senses that are visually distinct. For example, the keyword ``coach" can refer to multiple text semantics and visual senses (\textit{e.g.,} the ``bus", the ``handbag", the sports ``instructor", or the ``company"). This is commonly referred as word-sense disambiguation in Natural Language Processing \cite{wan2009latent}. 

Word-sense disambiguation is a top-down process arising from ambiguities in natural language. The text semantics of a word are robust and relatively static, and we can easily look them up from a dictionary resource such as WordNet \cite{miller1995wordnet} or Wikipedia. However, visual-sense disambiguation is a \textit{data-driven dynamic problem} which is specific to image collection. For the same keyword, the visual senses of images returned from the image search engine may be different at different time periods. For example, the keyword ``apple" might have mainly referred to the fruit before the company was founded.

Since the text semantics and visual senses of a given keyword are highly related, recent works also concentrated on combining text and image features \cite{chen2015sense,wan2009latent,loeff2006discriminating,yao2018aaai}. Most of these methods assume that there exists a one-to-one mapping between semantic and visual sense for a given keyword. However, this assumption is not always true in practice. For example, while there are two predominant text semantics of the word ``apple", there exist multiple visual senses due to appearance variation (green vs. red apples). To deal with the multiple visual senses, the method in \cite{chen2015sense} adopts a one-to-many mapping between text semantics and visual senses. This approach can help us discover multiple visual senses from the web but overly depends on the collected webpages. The effect of this approach will be greatly reduced if we can't collect webpages that contain enough text semantics and visual senses \cite{shen2018tmm,yao2018aaai}. 

Instead of relying on human-developed resources, we focus on automatically solving the visual disambiguation in an unsupervised way. Unlike the common unsupervised paradigm which jointly clusters text features and image features to solve the visual disambiguation, we present a novel framework that resolves the visual disambiguation by dynamically matching candidate text queries with retrieved images of the given keyword. Compared to human-developed and clustering-based methods, our approach can adapt to the dynamic changes in the search results. Our proposed framework includes two major steps: we first discover and dynamically select the text queries according to the keyword-based image search results, then we employ the proposed saliency-guided deep multi-instance learning (MIL) network to remove outliers and learn classification models for visual disambiguation. To verify the effectiveness of our proposed approach, we conduct extensive experiments on visual polysemy datasets CMU-Poly-30 and MIT-ISD to demonstrate the superiority of our approach. The main contributions are: 

1) Our proposed framework can adapt to the dynamic changes in search results and do visual disambiguation accordingly. Our approach has a better time adaptation ability.

2) We propose a saliency-guided deep MIL network to remove outliers and jointly learn the classification models for visual disambiguation. Compared to existing approaches, our proposed network has achieved the-state-of-the-art performance. 

3) Our work can be used as a pre-step before directly learning from the web, which helps to choose appropriate visual senses for sense-specific images collection, thereby improving the efficiency of learning from the web.

\section{The Proposed Approach}

\begin{figure}[tbp]
	\centering
	\includegraphics[width=0.5\textwidth]{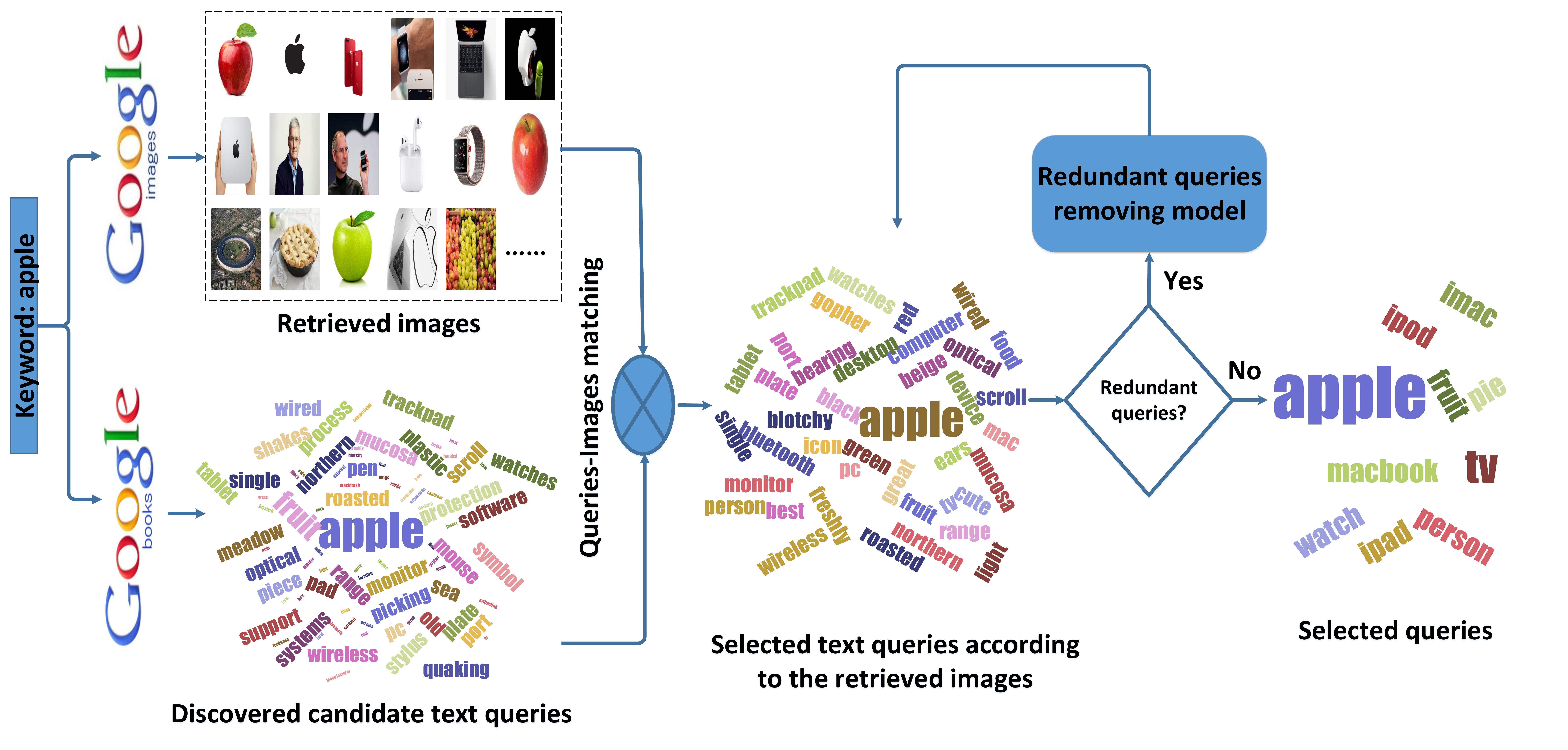}
	\caption{Framework of discovering and dynamically selecting text queries. The input is a keyword. We first discover a list of candidate text queries and retrieve the top images for the given keyword. Then we dynamically purify the candidate text queries according to the retrieved images. We remove the redundant and set the rest as selected text queries.}
	\label{fig2}  
\end{figure}

As shown in Fig. \ref{fig2} and Fig. \ref{fig3}, our proposed approach consists of two major steps. The following subsections describe the details of our proposed approach. 

\subsection{Discovering and Selecting Text Queries}

Inspired by recent works \cite{yao2018aaai,divvala2014learning}, untagged corpora Google Books \cite{lin2012syntactic} can be used to discover candidate text queries for modifying given keyword. Following the work in \cite{lin2012syntactic} (see section 4.3), we discover the candidate text queries by using n-gram dependencies whose modifiers are tagged as NOUN.

The image search results are dynamically changing, not all the candidate text queries have enough images in the search results representing their visual senses. Therefore, we can dynamically purify the candidate text queries by matching them with the retrieved images. 
Suppose the given keyword is $kw$, then we discover $E(kw)$ candidate text queries through Google Books. 
We collect the top $ K $ images for given keyword $ kw $. We perform a clean-up step for broken links and set the rest images $ I(kw) $ as the selected images for given keyword $ kw $ (\textit{e.g.,} ``apple"). In addition, we retrieve the top $ I(tq) = 5 $ images for each candidate text query $ tq $ (\textit{e.g.,} ``Apple laptop").
A text query $ tq \in E(kw) $ is expected to frequently appear in $ I(kw) $. To well obtain the visual senses of the images, some subset images which all have $ tq $ are required to contain visual similar content. To this end,  $ E(kw) $ can be selected in the following way.

For each image $ x_i \in I(tq) $, all images in $ I(kw) $ are matched with $ x_i $ on the basis of the visual similarity. In our work, the visual features and similarity measure methods in \cite{wang2014tpami} are leveraged. We set $ \vartheta _i(I) $ to be the number of images in $ I(kw) $ which can match with $ x_i $. The overall number of a candidate text query $ tq $ matching with the search results is its accumulated numbers over all the $ I(tq) $ images:
\begin{equation}\small\label{eq1}
\vartheta (tq) = \sum_{i=1}^{I(tq)}\vartheta_i(I). \\
\end{equation}
A large $ \vartheta (tq) $ indicates that $ tq $ matches in a good number of images in $ I(kw) $. When $ tq $ only presents in a few images or images involving $ tq $ are visually different, $ \vartheta (tq) $ will be set to be zero. Accordingly, when $ tq $ contains a big accumulated value $ \vartheta (tq) $, we can notice that lots of images within $ I(kw) $ contain $ tq $ and the images involving $ tq $ have similar visual senses. These $ N $ text queries with the highest numbers are chosen as the selected candidate text queries $ E(kw) $ for the given keyword $ kw $. 

Among the list of selected candidate text queries, some of them share visual similar distributions (\textit{e.g.,} ``Apple MacBook" and ``apple laptop"). To lessen the computing costs, these text queries which increase the discriminative power of the semantic space are kept and others are removed. 
To calculate the visual similarity between two text queries, half data in both text queries are used to learn a binary SVM classifier to do classification on the other half data. We believe that the two text queries are not similar if we can easily separate the testing data.
Assume we obtain $ N $ candidate text queries from the above step. We split the retrieved images of text query $ m $ into two groups, $ I_m^t $ and $ I_m^v $. To calculate the distinctness $ D(m,n) $ between $ m $ and $ n $, we train a binary SVM by using $ I_m^t $ and $ I_n^t $. 
We then obtain the probability of image in $ I_m^v $ belonging to the class $ m $ with the learned SVM classifier. Suppose the average score over $ I_m^v $ is $ \bar{\rho_m} $. Similarly, we can also obtain the average score $ \bar{\rho_n} $ over $ I_n^v $. Then $ D(m,n) $ can be calculated by:
\begin{equation}\small\label{eq2}
D(m,n)=\chi((\bar{\rho_m}+\bar{\rho_n})/2) \\
\end{equation}
where $ \chi $ is a monotonically increasing function. In this work, we define 
\begin{equation}\small\label{eq3}
\chi(\bar{\rho} )=1-e^{-\beta(\bar{\rho}-\alpha ) } \\
\end{equation}
in which the parameters $ \alpha $ and $\beta$ are two constants. When the value of $(\bar{\rho_m}+\bar{\rho_n})/2$ goes below the threshold $ \alpha $, $ \chi(\bar{\rho} ) $ decreases with a fast speed to penalize pair-wisely similar text queries. In our work, the value of $ \alpha $ and $\beta$ are set to be 0.6 and 30 respectively. 

Finally, we select a set of text queries from the $ N $ candidates. The selected text queries are most relevant to the given keyword $ kw $. We define the relevance in \eqref{eq1}. Meanwhile, to characterize the visual distributions of the given keyword, the selected text queries are required to dissimilar with each other from the visual relevance perspective.
The distinctiveness can be calculated through matrix $ D $ in \eqref{eq2}. We can solve the following optimization problem to satisfy the two criteria.

$ \gamma $ is used to indicate text query $ n $ is selected or removed. Specifically, we set $ \gamma_n = 1 $ when selected and $ \gamma_n = 0 $ when removed. We can estimate the value of $ \gamma $ by solving:
\begin{equation}\small\label{eq3-2}
\arg \max_{\gamma\in\{0,1\}^N}\{\lambda \phi_\gamma+\gamma^{N}D_\gamma \} \\
\end{equation}
Let $ {tq}_n $ be the text query of keyword $ kw $. $ \phi =(\vartheta ({tq}_1),\vartheta ({tq}_2),...,\vartheta ({tq}_N)) $, where $ \vartheta ({tq}_n) $ is defined in \eqref{eq1}. $ \lambda $ is the scaling factor. Due to the integer quadratic programming is NP hard, $\gamma$ is relaxed to be in $ \mathbb{R}^T $ and we choose text query $ n $ whose $ \gamma _n \geqslant 0.5 $ as the final selected text query.

\subsection{Saliency-guided Deep MIL Model}

\begin{figure}[tbp]
	\centering
	\includegraphics[width=0.5\textwidth]{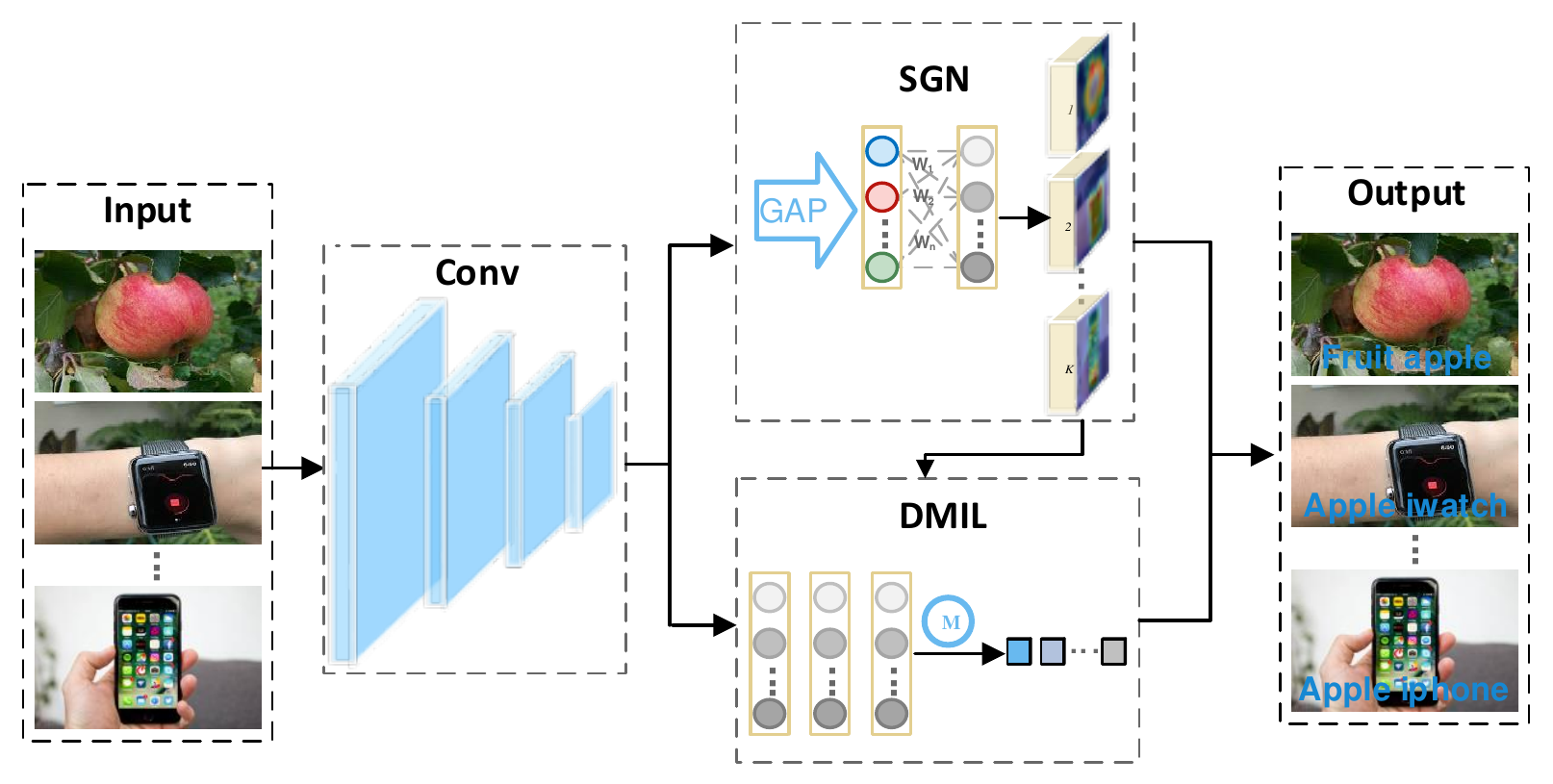}
	\caption{The framework of saliency-guided deep MIL model. Our framework includes two stream networks SGN and DMIL. SGN is to localize object and generate an instance for the web image. DMIL is to encode the discriminative features for learning the deep MIL classification models to remove outliers and perform visual disambiguation.}
	\label{fig3} 
\end{figure}

Due to the error indexing of image search engine, even we retrieve the top few sense-specific images, some noise may still be included \cite{shen2018tmm,yao2018aaai}. As shown in Fig. \ref{fig3}, our model consists of two stream networks SGN and DMIL. SGN is to localize object for generating instance. DMIL is to encode the discriminative features for learning the deep classification models to remove outliers and perform visual disambiguation. 

Different from existing methods which attempt to follow a multi-instance assumption, where its object proposals can be regarded as one ``instance" sets and each image can be treated as one ``bag", our approach treats each selected text query as a ``bag" and each image therein as one ``instance". The main reason for this is that our images come from the web and may contain noise. If we treat each web image as a ``bag", the generated proposals (``instances") by existing methods like selective search \cite{selective2013} or RPN \cite{rpn2015} can't always satisfy such a condition: object lies in at least one of the proposals. However, when we treat each image returned from the image search engine as one ``instance", and each selected text query as one ``bag", then it becomes natural to formulate outliers removal as a multi-instance learning problem.

The selected text queries are leveraged to collect sense-specific images from the image search engine. To reduce the interference of noisy background objects in web images, we propose to use a saliency extraction network (SGN) to localize the discriminative regions and generate the instance for the web image. Specifically, we follow the work in \cite{saliency2016} to model this process by leveraging global average pooling (GAP) to produce the saliency map. The feature maps of the last convolutional layer with weights were summed to generate the saliency map for each image. Finally, we conduct binarization operation on the saliency map with a adaptive threshold, which is obtained through OTSU algorithm \cite{threshold1979}. We leverage the bounding box that covers the largest connected area as the discriminative region of object. For a given image \textit{I}, the value of spatial location (\textit{x}, \textit{y}) in saliency map for category \textit{c} is defined as follows:
\begin{equation}\small\label{eq4}
\textit{M}_{c} (\textit{x},\textit{y})= \sum_{\textit{u}} \textit{w}_{\textit{u}}^{c}f_{\textit{u}}(\textit{x}, \textit{y}) \\
\end{equation}
where $\textit{M}_{c} (\textit{x}, \textit{y})$ directly indicates the importance of activation at spatial location (\textit{x}, \textit{y}) leading to the classification of an image to category \textit{c}. $f_{\textit{u}}(\textit{x}, \textit{y})$ denotes the activation of neuron \textit{u} in the last convolutional layer at spatial location (\textit{x}, \textit{y}), and $\textit{w}_{\textit{u}}^{c}$ denotes the weight that corresponding to category \textit{c} for neuron \textit{u}. Instead of treating the whole image as one instance, we use the generated bounding box result as the instance.

In traditional supervised learning paradigm, training samples are given as pairs $\{(\textit{x}_i, \textit{y}_i)\}$, where $\textit{x}_i \in \mathbb{R}^{d}$ is a feature vector and $\textit{y}_i \in \{-1, 1\}$ is the label. However, in MIL, data are organized as bags $\{\mathbf{X}_{i}\}$. Each bag contains a number of instances $\{\textit{x}_{i,j}\}$. Labels $\{\mathbf{Y}_i\}$ are only available for the bag. The labels of instances $\{\textit{y}_{i,j}\}$ are unknown.
Considering the recent advances achieved by deep learning, in this work, we propose to exploit deep CNN as our architecture for learning visual representation with multi-instance learning. Our structure is based on VGG-16 \cite{simonyan2014very} and we redesign the last hidden layer for MIL.
For a given training image $ \textit{x} $, we set the output of the last fully connected layer $fc_{15} \in \mathbb{R}^{m}$ as high-level features of the input image.
Followed by a softmax layer, $ fc_{15} $ is transformed into a probability distribution $\rho \in \mathbb{R}^{m}$ for objects belonging to the $m$ text queries. Cross-entropy is taken to measure the prediction loss of the network.
Specifically, we have
\begin{equation}\small\label{eq5}
L=-\sum_{i}t_{i}\log(\rho_i) \:\:\: \text{where} \:\:\: \rho_{i}=\frac{\exp(h_i)}{\sum_{i}\exp(h_i) }, i = 1..,m. \\
\end{equation}
We can calculate the gradients of the deep CNN through back-propagation
\begin{equation}\small\label{eq6}
\frac{\partial L}{\partial h_i}=\rho _i - t_i, \\
\end{equation}
where 
\begin{equation}\small\label{eq7}
t = \{t_i| \sum_{i=1}^{m}t_i = 1, \:\: t_i \in \left \{ 0,1 \right \}, i = 1,...,m \}
\end{equation}
represents the true label of the sample \textit{x}.
To learn multiple instances as a bag of samples, we incorporate deep representation with MIL and name it as DMIL. 
Assume a bag $\{\textit{x}_j|j=1,...,n\}$ contains $ n $ instances and the label of the bag is $t=\{t_i|t_i \in \{0,1\}, i = 1,...m\}$; DMIL extracts representations of the bag: $h=\{h_{ij}\} \in R^{m \times n}$, in which each column is the representation of an instance. The aggregated representation of the bag for MIL is:
\begin{equation}\small\label{eq8}
\tilde{h_i} = f(h_{i1},...,h_{in})
\end{equation}
where function $ f $ can be $\max_j (h_{ij})$, $\text{avg}_j (h_{ij})$, or $\log [1+\sum_{j} \exp(h_{ij})]$. For this work, we use the $ \max(\cdot ) $ layer. In the ablation studies, we show experiments with these possible choices.
Then we can represent the visual distribution of the bag and the loss \textit{L} as:
\begin{equation}\small\label{eq9}
\rho_{i}=\frac{\exp({\tilde{h_i}})}{\sum_{i}\exp({\tilde{h_i}})}, i = 1,...,m.  
\end{equation}
and 
\begin{equation}\small\label{eq10}
L=-\sum_{i}t_{i}\log(\rho_i)
\end{equation}
respectively. To minimize the loss function of DMIL, we employ stochastic gradient descent (SGD) for optimization. The gradient can be calculated via back propagation \cite{back1986}:
\begin{equation}\small\label{eq11}
\frac{\partial L}{\partial \tilde{h_i}}=\rho _i - t_i \:\: \text{and} \:\: \frac{\partial \tilde{h_i}}{\partial h_{ij}}=\left\{\begin{matrix}
1, & h_{ij}= \tilde{h_i}\\ 
0, & \text{else}
\end{matrix}.\right.
\end{equation}

For the task of disambiguating the keyword-based image search results, we first employ SGN to generate the saliency map for localizing the discriminative region and generating the ``instance" of the image. Then the proposed DMIL is to encode the discriminative features for learning deep models to remove outliers and perform visual disambiguation.

\section{Visual Disambiguation Experiments}

\subsection{Datasets and Evaluation Metric}

Two widely used polysemy datasets CMU-Polysemy-30 \cite{chen2015sense} and MIT-ISD \cite{saenko2009unsupervised} are employed to validate the proposed framework. Specifically, we set the images corresponding to various keywords in CMU-Poly-30 and MIT-ISD as the results of keyword-based image search. We follow the setting in baselines \cite{chen2015sense,yao2018aaai} and exploit web images as the training set, human-labeled images in CMU-Polysemy-30 and MIT-ISD as the testing set. Average Classification Accuracy (\textbf{ACA}) is adopted as the evaluation metric. If there is no special statement, the image features are 4096-dimensional deep features based on VGG-16 model.

\begin{figure}[tbp]
	\centering
	\includegraphics[width=0.46\textwidth]{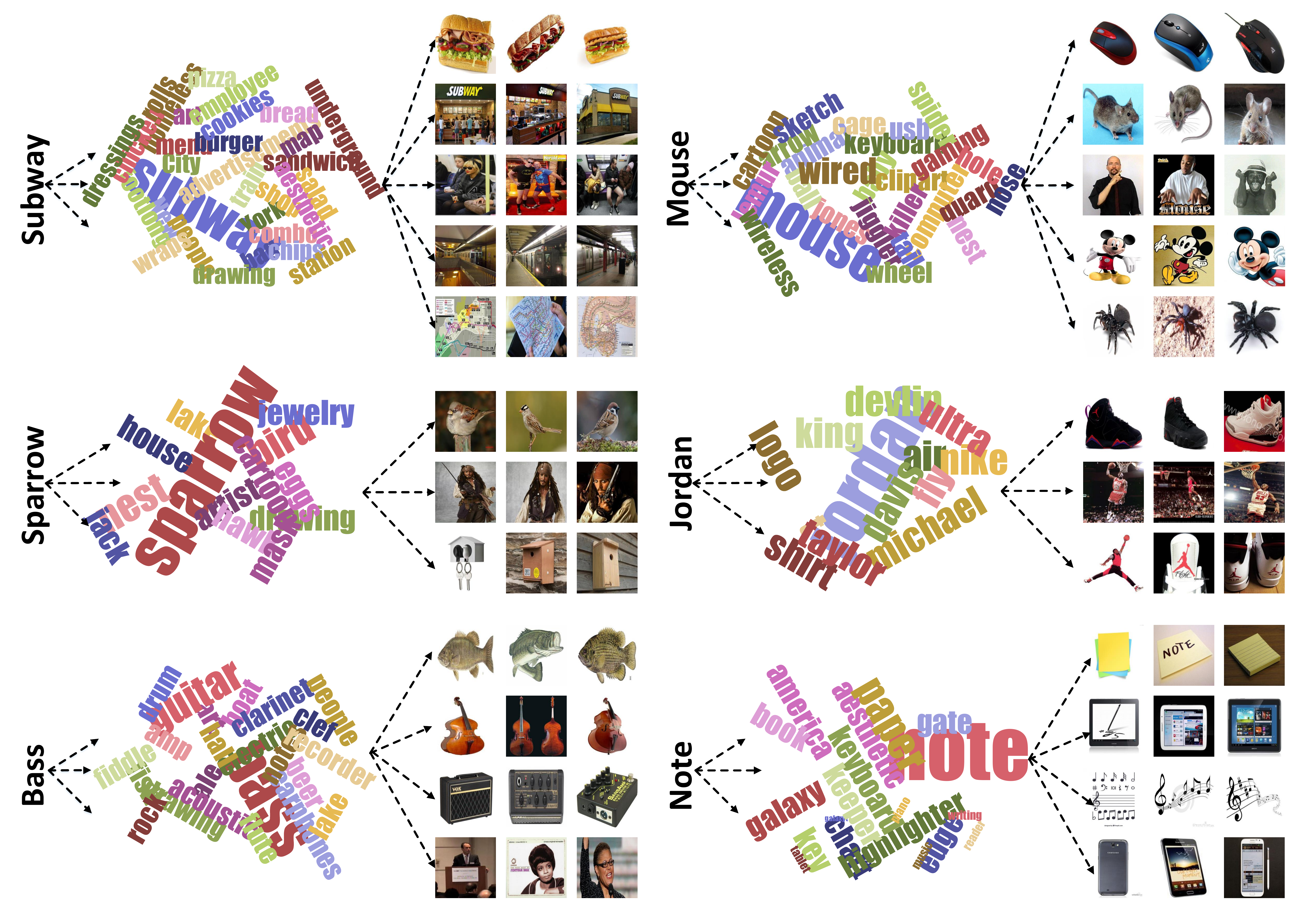}
	\caption{A snapshot of multiple text queries discovered from Google Books and visual senses disambiguated from the CMU-Poly-30 dataset by our proposed framework. For example, our proposed method automatically discovers and disambiguates five senses for ``$ Subway $": subway sandwich, subway store, subway people, subway station and subway map. For ``$ Mouse $", it discovers multiple visual senses of the computer mouse, mouse animal, mouse man, and cartoon mouse $ etc $.}
	\label{fig4} 
\end{figure}

\subsection{Implementation Details and Parameters}

For each keyword, we first discover the candidate text queries by searching in the Google Books. We set the corresponding images in CMU-Polysemy-30 and MIT-ISD as the results of keyword-based image search. Then we retrieve the top $ I(tq) $ images for each candidate text query. The value of $ I(tq) $ is selected from $ \{1,2,3,4,5,6,7,8,9\} $. We dynamically purify the candidate text queries by matching them with the results of keyword-based image search. Specifically, we select the top $ N $ text queries with the highest numbers. $ N $ is selected from $ \{10,20,30,40,50,60\} $. For removing redundancy and selecting representative text queries, we retrieve the top 100 images for the selected candidate text queries and assume the retrieved images are positive instances (in spite of the fact that noisy images might be included). The collected 100 images for each selected text query were randomly split into a training set and testing set (\textit{e.g.,} $I_m=\{I_m^t = 50, I_m^v = 50\}$ and $I_n=\{I_n^t = 50, I_n^v = 50\}$). We train a linear SVM classifier with $ I_m^t $ and $ I_n^t $ for classifying $ I_m^v $ and $ I_n^v $ to obtain the value of $ \bar{\rho_m} $ and $ \bar{\rho_n} $. We then get the distinctness $ D(m,n) $ by calculating \eqref{eq2} and remove redundant queries by solving \eqref{eq3-2}. $ \alpha $ is selected from $ \{0.2,0.4,0.5,0.6,0.8\} $ and $\beta$ is selected from $ \{10,20,30,40,50\} $ in \eqref{eq2}. $ \gamma_n$ is set $ \gamma_n\geqslant 0.5 $ in \eqref{eq3-2}.

The structure of SGN is based on VGG-16 \cite{simonyan2014very}. To obtain a higher spatial resolution, we remove the layers after conv$ 5\_3 $ and get a mapping resolution of 14 $ \times $ 14. Then we add a convolutional layer of size 3 $ \times $ 3, stride 1, pad 1 with 1024 neurons, followed by a global average pooling (GAP) layer and a softmax layer. SGN is pre-trained on the 1.3 million images of ImageNet dataset \cite{deng2009imagenet} and then fine-tuned on the collected web images. The number of neurons in the softmax layer is set as the number of selected text queries. The structure of DMIL is also based on VGG-16 \cite{simonyan2014very}. We remove the last hidden layer and use the $ \max(\cdot ) $ layer instead. The initial parameters of the modified version of the model are inherited from the pre-trained VGG-16 model. During training, we leveraged ``instances" generated by SGN and set the selected text queries as ``bags" to fine-tune the model. DMIL is trained for 100 epochs with an initial learning rate selected from [0.0001, 0.002] (which is robust). In order to generate test ``bags", we solely sampled images from the CMU-Polysemy-30 and MIT-ISD datasets. 

\begin{table}[tb]
	\centering
	\renewcommand{\arraystretch}{1.0}\small
	\caption{Visual disambiguation results (ACA) on two evaluated datasets CMU-Poly-30 and MIT-ISD. The best result is marked in bold.}
	\begin{tabular}{c|c|c|c}
		\hline
		\multirow{2}{*}{}        &  \multirow{2}{*}{\textbf{Method}}    & \multicolumn{2}{c}{\textbf{Dataset}} \\      
		\cline{3-4}              &                                      &  CMU-Poly-30          &   MIT-ISD        \\
		\hline
		\multirow{6}{*}{\S}      & VSD \cite{wan2009latent}             & 0.728                    &   0.786         \\
		& ULVSM \cite{saenko2009unsupervised}  & 0.772                    &   0.803         \\    
		& WSDP \cite{barnard2005word}          & 0.791                 &   0.743         \\
		& NEIL \cite{chen2013neil}             & 0.741                 &   0.705         \\
		& ConceptMap \cite{golge2014}          & 0.726                 &   0.758         \\
		& VSCN \cite{qiu2013}                  & 0.802                 &   0.783         \\
		\hline
		\multirow{7}{*}{\$}      & ISD \cite{loeff2006discriminating}   & 0.554                    &   0.634         \\
		& IWSD \cite{lucchi2012}               & 0.643                    &   0.725         \\
		& SDCIT \cite{chen2015sense}           & 0.839                    &   0.853         \\
		& VSDE \cite{acl2016}                  & 0.747                 &   0.763         \\                         
		& LEAN \cite{divvala2014learning}      & 0.827                 &   0.814         \\
		& DRID \cite{zhang2016domain}            & 0.846                 &   0.805         \\
		& DDPW \cite{shen2018tmm}               & 0.884                 &   0.897         \\
		\hline
		\multirow{1}{*}{\P}      & SG-DMIL (Ours)                        & \textbf{0.925}        &   \textbf{0.938}\\             
		\hline
	\end{tabular}\\
	\vspace{0.1cm}
	\leftline{\qquad\: \$ \: : combination of text and image based methods}
	\leftline{\qquad\: \S \:\: : image-based methods \P \: : our proposed approach}
	\label{tab1}
\end{table} 

\subsection{Baselines}

The method in \cite{shen2018tmm} reproduced nearly all leading methods on the CMU-Polysemy-30 and MIT-ISD dataset. Specifically, we compare our approach with two groups of baselines: image-based methods and the combination of text and image based methods. 

\subsection{Results and Analysis}

Fig. \ref{fig4} shows a snapshot of multiple text queries discovered from Google Books and visual senses disambiguated from the CMU-Poly-30 dataset by our proposed framework. It should be noted that for some keywords, CMU-Poly-30 dataset only annotates one or two visual senses. However, our proposed approach successfully discovers and distinguishes more visual senses. For example, for keyword ``bass" in CMU-Poly-30 dataset, only ``bass fish" and ``bass guitar" are annotated. Our approach additionally discovered two other visual senses ``bass amp" and ``Mr./Miss Bass". This is mainly due to our approach can dynamically select text queries based on image search results. 

To leverage the ground truth labels in CMU-Poly-30 and fairly compare with other baseline methods, we remove the text queries discovering and selecting procedure and directly use the annotated labels in the dataset to collect web images. Then we leverage the proposed saliency-guided deep MIL to remove outliers and train classification models for visual disambiguation. Table \ref{tab1} presents the ACA results on the CMU-Poly-30 and MIT-ISD dataset.
By observing Table \ref{tab1}, our proposed approach achieves the-state-of-the-art ACA performance on CMU-Poly-30 and MIT-ISD dataset, which produces significant improvements over image-based methods, and the combination of text and image based methods. One possible explanation is that our proposed saliency-guided deep MIL can effectively remove the outlier images from the image search results and train robust classification models for visual disambiguation. 

\section{Ablation Studies}

\subsection{Coefficients in Proposed Framework}

\begin{figure}[h]
	\centering
	\includegraphics[width=0.46\textwidth]{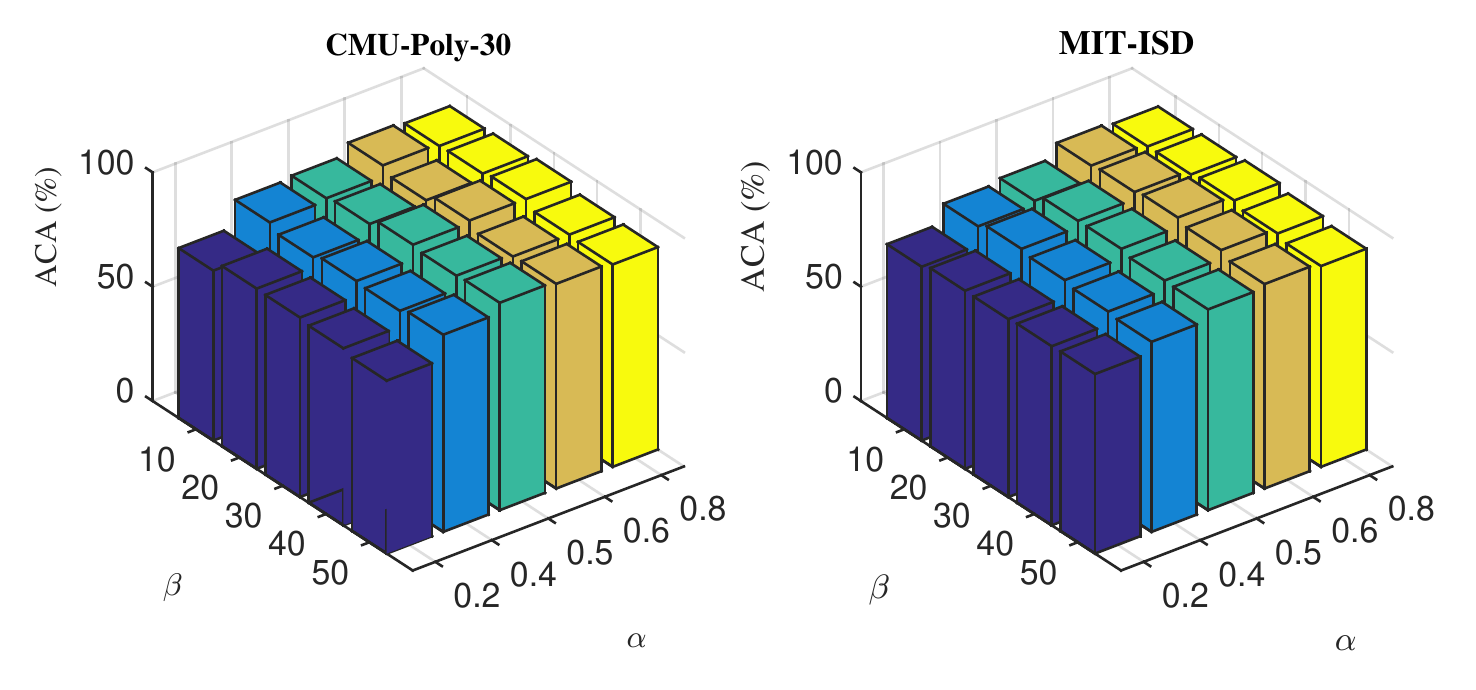}
	\caption{The ACA performance of the interaction between pairs of parameters $\alpha $ and $\beta$.}
	\label{fig5} 
\end{figure}

\begin{figure}[h]
	\centering 
	\begin{minipage}[b]{0.235\textwidth} 
		\centering 
		\includegraphics[width=0.95\textwidth]{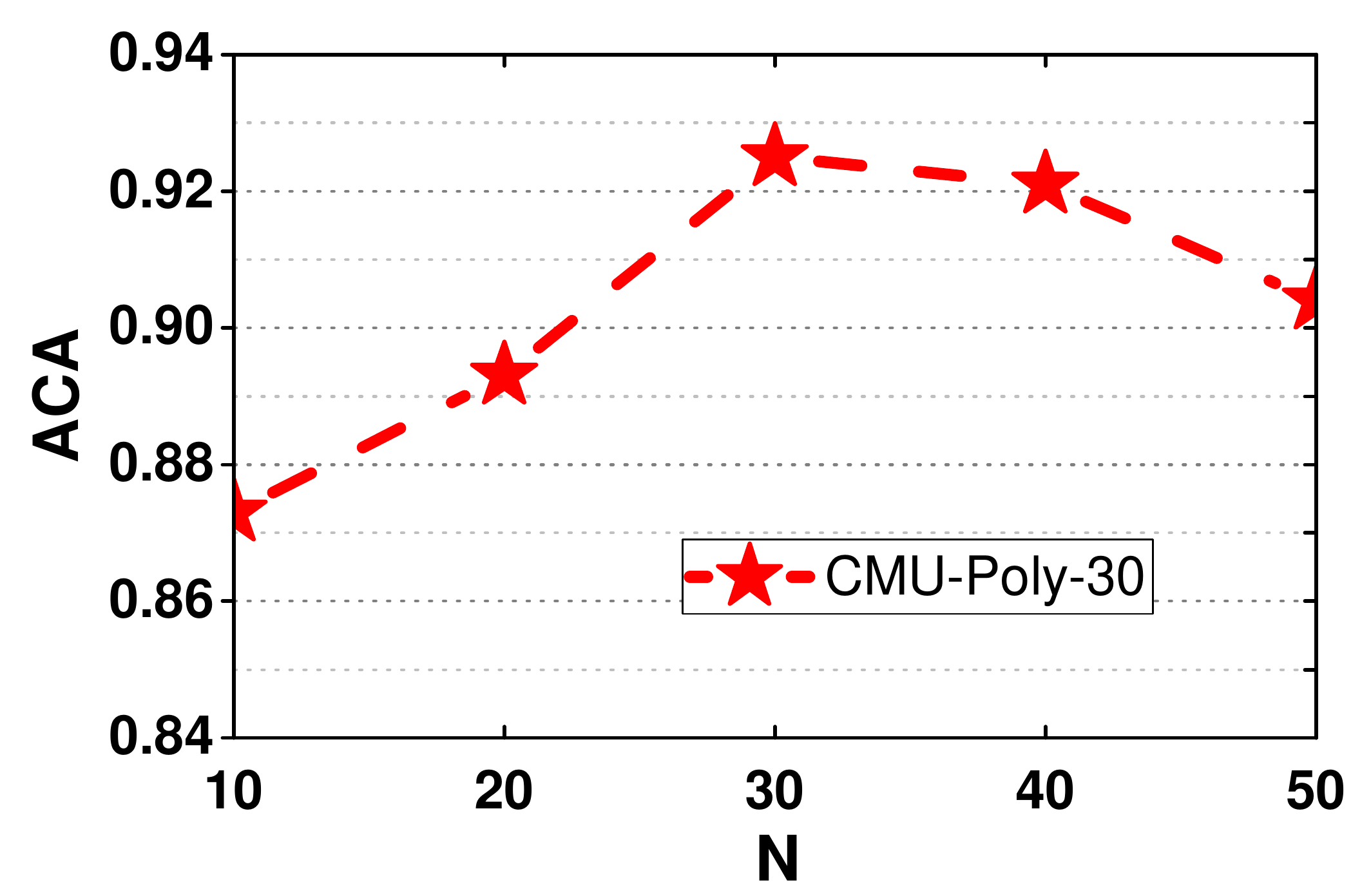} 
	\end{minipage}
	\begin{minipage}[b]{0.235\textwidth} 
		\centering 
		\includegraphics[width=0.95\textwidth]{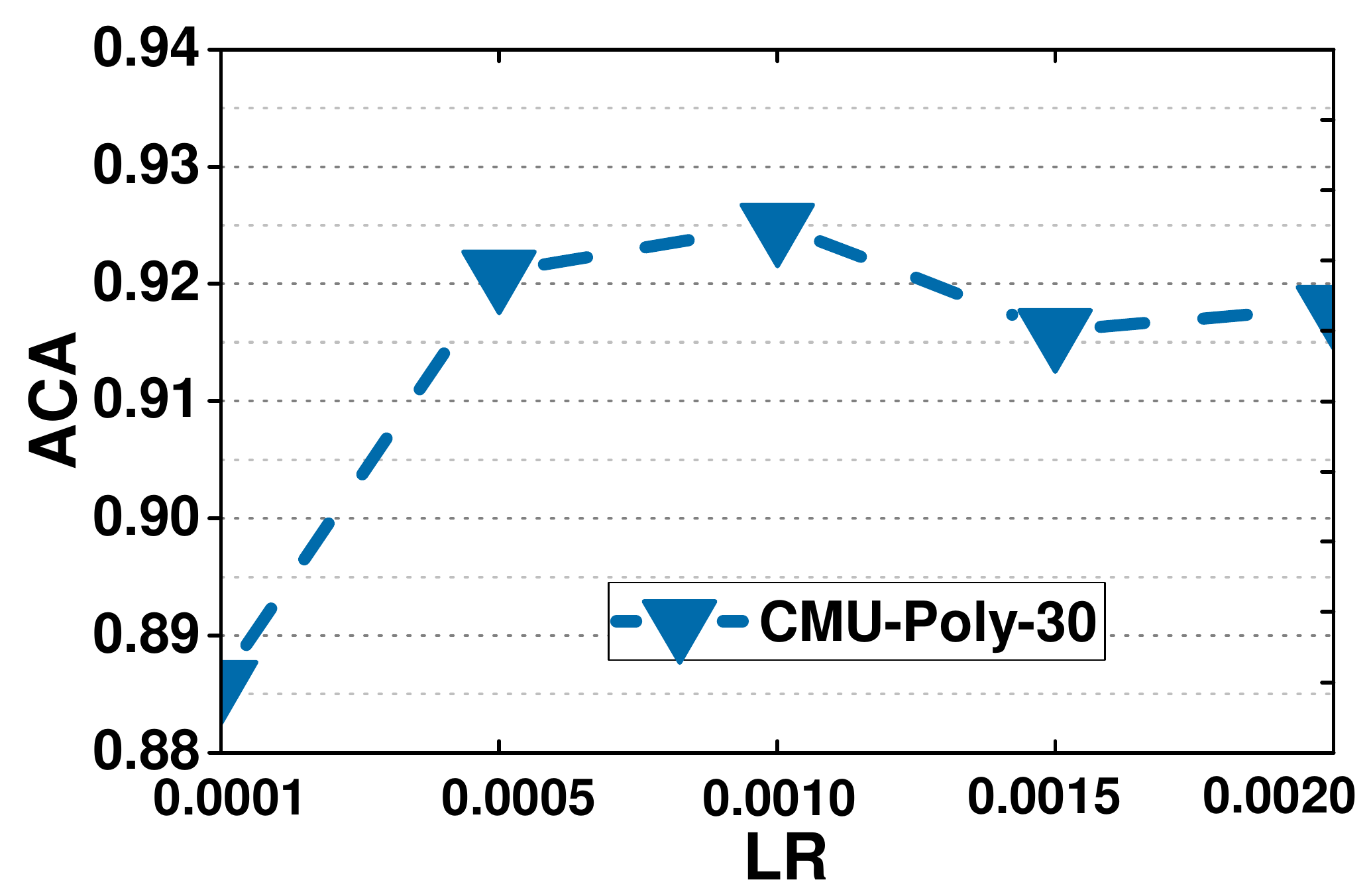}
	\end{minipage}\\
	\begin{minipage}[b]{0.235\textwidth} 
		\centering 
		\includegraphics[width=0.95\textwidth]{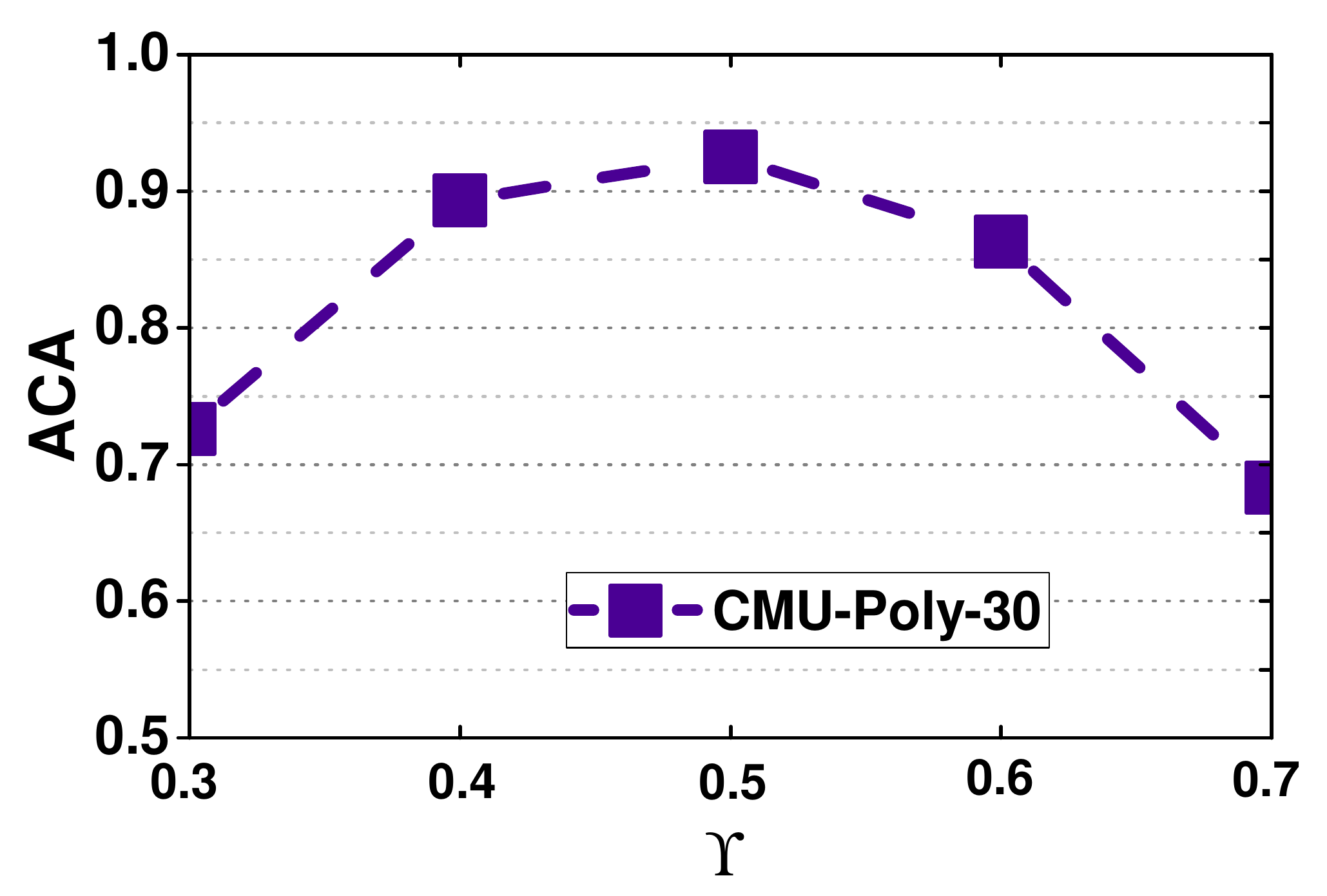} 
	\end{minipage}
	\begin{minipage}[b]{0.235\textwidth} 
		\centering 
		\includegraphics[width=0.95\textwidth]{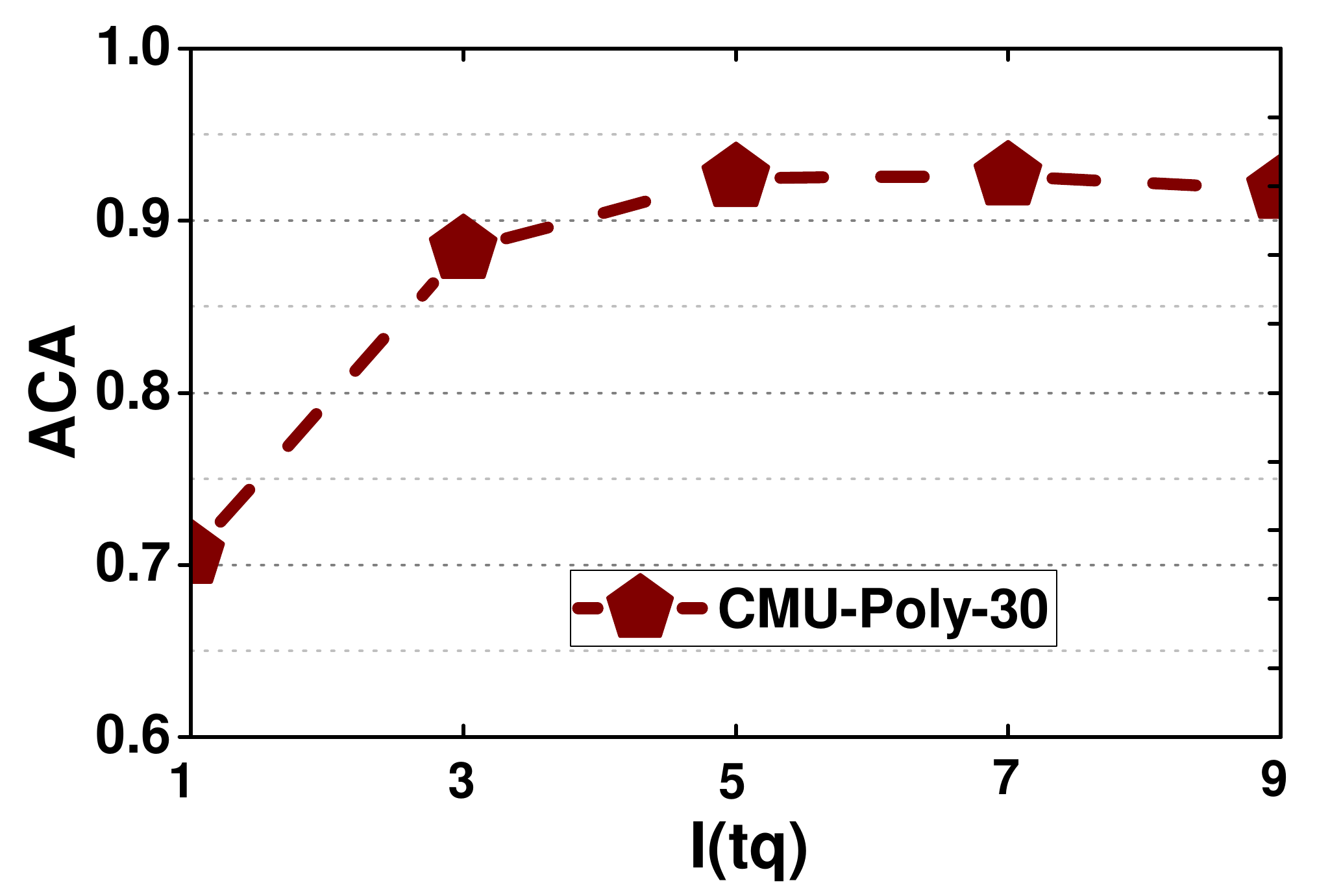}
	\end{minipage}
	\caption{The parameter sensitivities of $N$, LR, $\gamma$, and $ I(tq) $ w.r.t. ACA in CMU-Poly-30 dataset.}
	\label{fig6}
\end{figure}

For the coefficients analysis, we mainly concern the parameters $\alpha $, $\beta$, $\gamma$, $N$, and $ I(tq) $ in selecting text queries and learning rate (LR) in DMIL. Specifically, we analyze the interaction between pairs of parameters $\alpha $ and $\beta$ in Eq \eqref{eq3}. For other parameters, we analyze the sensitivities by a graphic per parameter. As shown in Fig. \ref{fig5}, the changing tendency of ACA w.r.t. ($\alpha, \beta$), overall, is stable and consistent. Fig. \ref{fig6} presents the parameter sensitivities of $N$, LR, $\gamma$, and $ I(tq) $ w.r.t. ACA in CMU-Poly-30 dataset.

\subsection{Pre-step before Learning from the Web}

Our work can be used as a pre-step before directly learning from the web. To verify this statement, we collected the top 100 web images from the Google Image Search Engine by using the labels in CUB-200-2011 dataset \cite{branson}. Then we employed the proposed approach to choose appropriate visual senses and purify the outliers. The outputs are a set of relatively clean web images. We leveraged the relatively clean web images as the training set to perform one of the most popular weakly supervised fine-grained algorithms Bilinear \cite{bilinear} on CUB-200-2011 \cite{branson} testing set. The results are shown in Table \ref{tab2}. By observing Table \ref{tab2}, we can observe that our proposed approach greatly improves the baseline accuracy. 

\begin{table}[t]
	\centering
	\renewcommand{\arraystretch}{1.0}\small
	\caption{Fine-grained visual recognition results on CUB-200-2011 testing set.}
	\begin{tabular}{c|c|c}
		\hline
		Training data                      & Algorithm    & Accuracy  \\
		\hline
		Original web                       & Bilinear     & 0.718     \\
		\textbf{Clean web}                   & Bilinear   & 0.752     \\
		CUB training                       & Bilinear     & 0.841     \\
		\textbf{Clean web + CUB training}  & Bilinear     & 0.863     \\
		\hline
	\end{tabular}
	\label{tab2}
\end{table} 

\begin{figure} [t]
	\centering
	\subfloat[]{
		\includegraphics[width=0.22\textwidth]{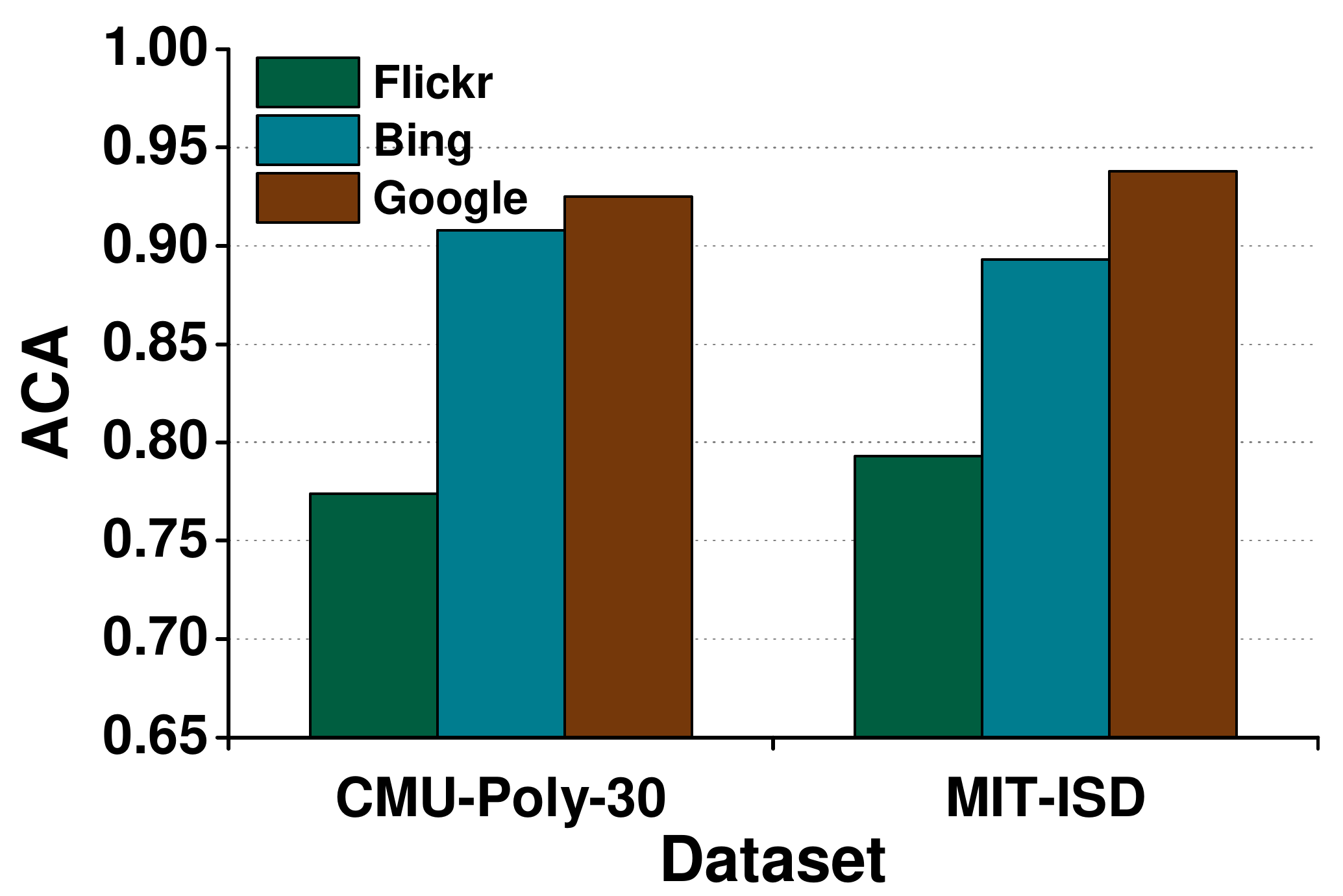}}
	\subfloat[]{
		\includegraphics[width=0.22\textwidth]{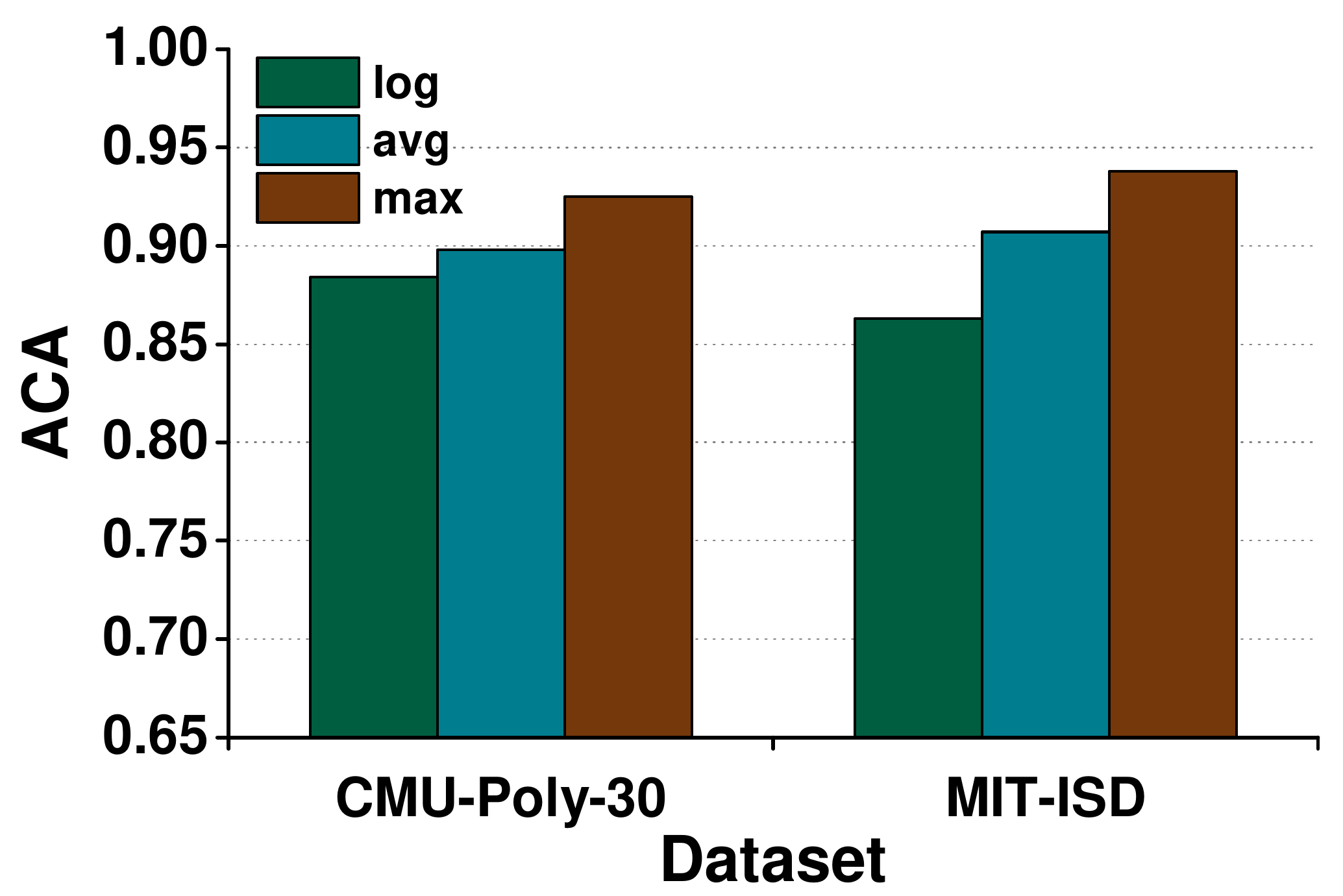}}
	\caption{(a) Impact of different domains. (b) Impact of different hidden layers.}
	\label{fig8}
\end{figure}

\subsection{Influence of Different Domains}

To analyze the influence of using different domain web images for visual disambiguation, we collected web images for selected text queries from the Google Image Search Engine, the Bing Image Search Engine, and Flickr respectively. As shown in Fig \ref{fig8} (a),  the performance of web images coming from Flickr is much lower than from the Google Image Search Engine and the Bing Image Search Engine. The performance of web images coming from the Google Image Search Engine is a little better than from the Bing Image Search Engine. 

\subsection{Influence of Different Hidden Layer}

The choice of hidden layer is of critical importance in our proposed saliency-guided deep MIL network. As mentioned in Section 3.2, the $ \max(\cdot ) $, $ \text{avg}(\cdot ) $, and $ \log(\cdot ) $ refer to $\max_j (h_{ij})$, $\text{avg}_j (h_{ij})$, and $\log [1+\sum_{j} \exp(h_{ij})]$, respectively. From Fig \ref{fig8} (b), we can notice that the straightforward $ \max(\cdot) $ layer obtains the best ACA performance.

\begin{figure}[t]
	\centering
	\subfloat[]{\includegraphics[width=0.22\textwidth]{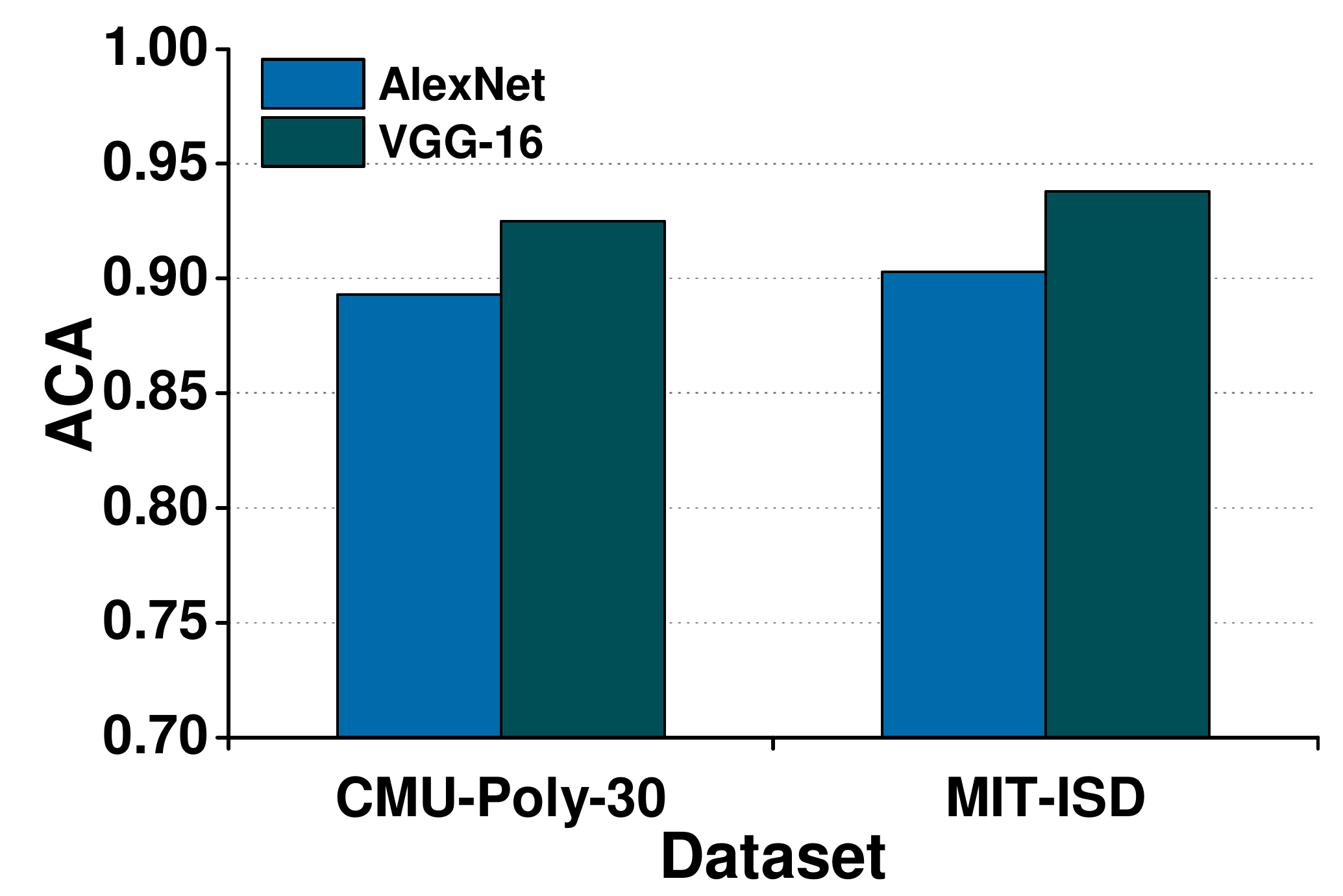}}
	\subfloat[]{\includegraphics[width=0.22\textwidth]{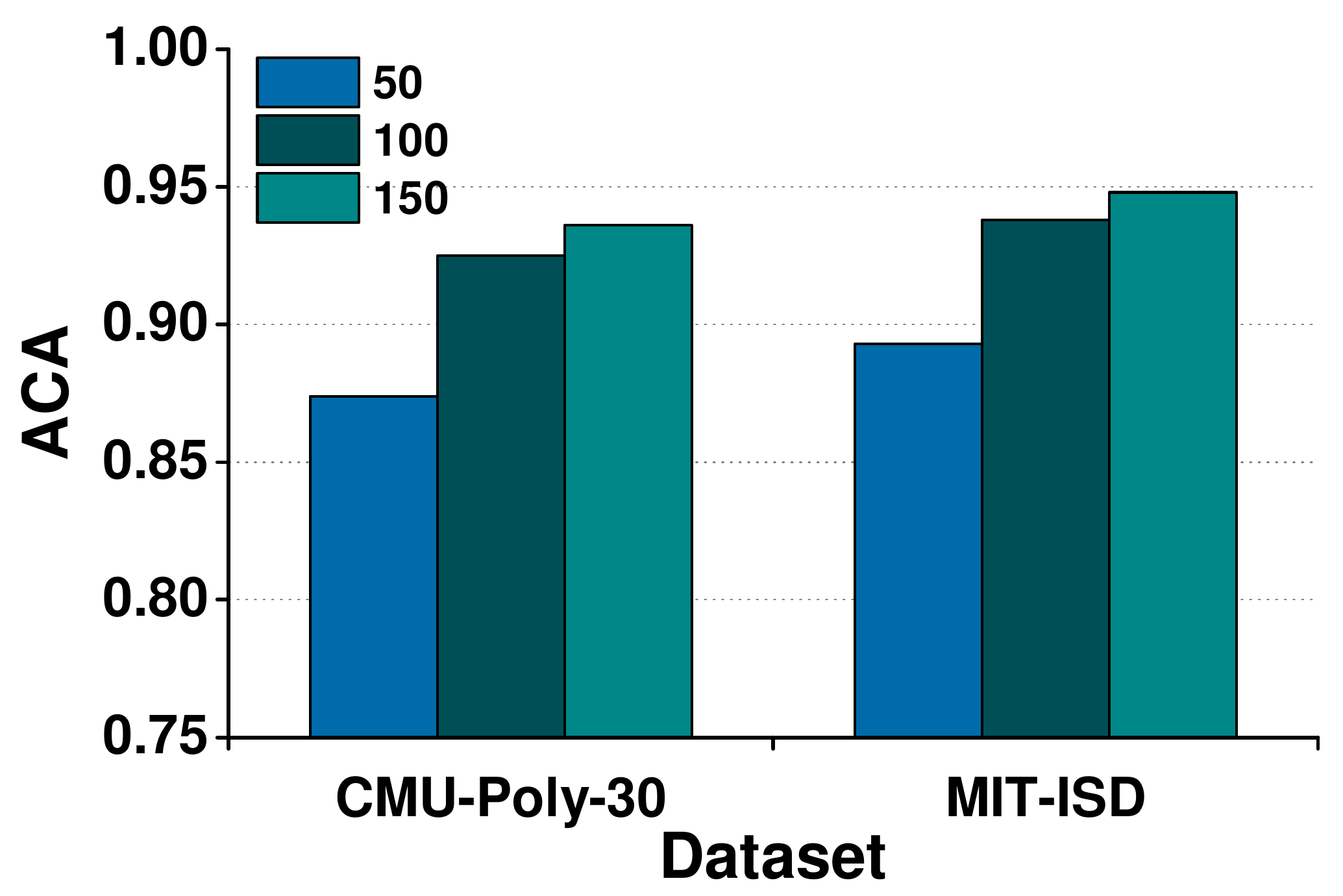}}
	\caption{(a) Impact of different CNN architectures. (b) Impact of different training samples.}
	\label{fig7}
\end{figure}

\subsection{Are Deeper Models Helpful?}

It is well known that the CNN model architecture has a critical impact on object recognition performance. We investigated this issue by replacing VGG-16 with a shallower architecture AlexNet in the saliency-guided deep MIL model and compared the results. As shown in Fig \ref{fig7} (a), using a deeper model (VGG-16) was better than using shallower models (AlexNet), as expected. In particular, the VGG model was more effective for localizing the objects from the images. 

\subsection{Are More Web Images Helpful?}

Data scale has a large impact on web-supervised learning. We investigated this impact by incrementally increasing or decreasing the number of web images used for each text query. Specifically, we choose $\{50,100,150\}$ images from the web for each selected text query. As shown in Fig \ref{fig7} (b), in general, the performance of ACA improved steadily with the use of more training samples.

\begin{figure}[h]
	\centering
	\includegraphics[width=0.48\textwidth]{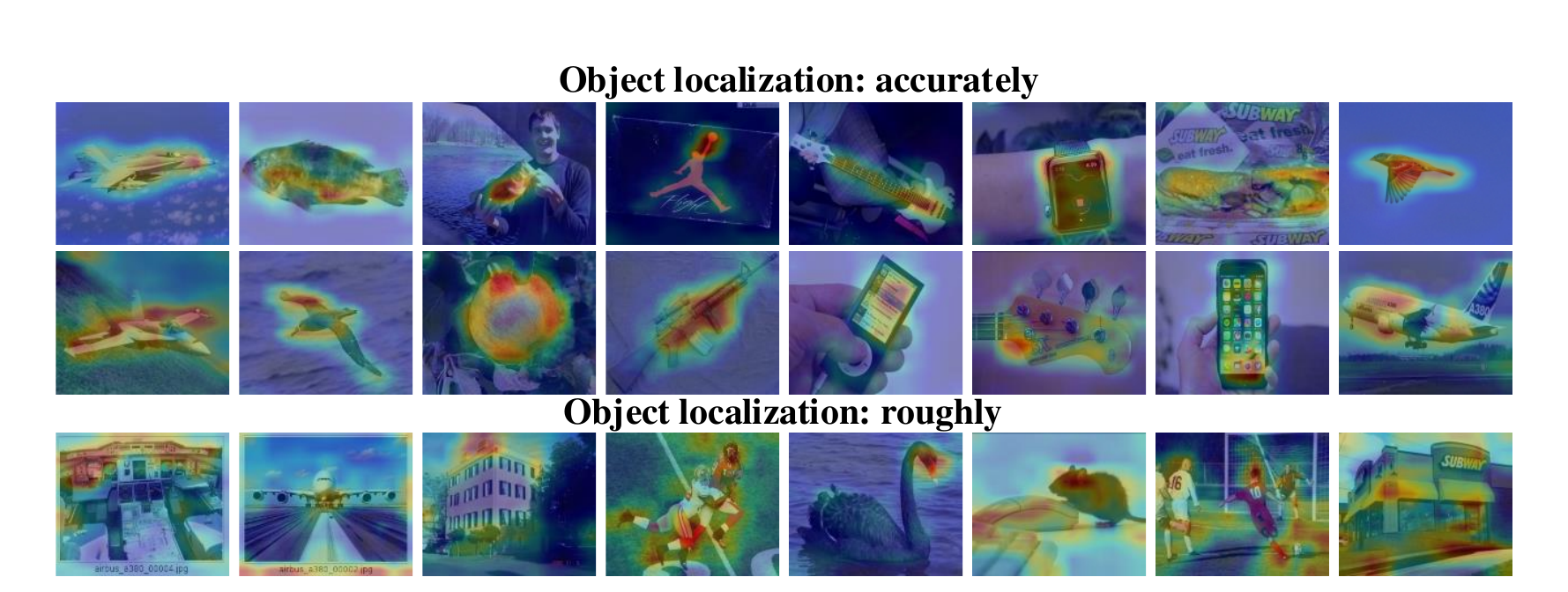}
	\caption{Visualization of object locating via saliency map.}
	\label{fig9}  
\end{figure}
\subsection{Visualization}

Our saliency-guided deep MIL model consists of two stream networks, and SGN is to localize object and generate the ``instance" for the web images. Whether or not accurately locate the object for the DMIL network is the basis for extracting deep features and learning the classification models. Fig. \ref{fig9} visualizes the object locating via saliency map. By observing Fig. \ref{fig9}, we can find the SGN can well locate the object for the web image. For some images, although SGN cannot accurately locate the exact location where it is located, the rough region of SGN locating still contains the location of the object.  

\section{Conclusions}

In this work, we focused on one important yet often ignored problem: we argue that the current poor performance of some models learned from the web images is due to the inherent ambiguity in user queries. We solved this problem by visual disambiguation in search results. Our work could be used as a pre-step before directly learning from the web images, which helps to choose appropriate visual senses for images collection and thereby improve the efficiency of learning from the web. Compared to existing methods, our proposed approach can 1) figure out the right visual senses, and 2) adapt to the dynamic changes in the search results. Extensive experiments demonstrated the superiority of our proposed approach.

\section*{Acknowledgments}

This work was supported in part by the National Natural Science Foundation of China under Project 61502081, Sichuan Science and Technology Program (No.2018GZDZX0032),


\begin{thebibliography}{1}\itemsep=-1pt
	
	\bibitem[\protect\citeauthoryear{Simonyan}{2007}]{simonyan2014very}
	K Simonyan et al,
	\newblock Very deep convolutional networks for large-scale image recognition.
	\newblock \textit{arXiv:1409.1556}, 2014.
	
	\bibitem[\protect\citeauthoryear{Deng}{2009}]{deng2009imagenet}
	J~Deng et al,
	\newblock Imagenet: A large-scale hierarchical image database.
	\newblock \textit{CVPR}, 2009.
	
	\bibitem[\protect\citeauthoryear{Yao}{2018}]{yao2018aaai}
	Y~Yao et al,
	\newblock Discovering and Distinguishing Multiple Visual Senses for Polysemous Words.
	\newblock \textit{AAAI}, 2018.
	
	\bibitem[\protect\citeauthoryear{Shen}{2019}]{shen2018tmm}
	F Shen et al,
	\newblock Extracting Multiple Visual Senses for Web Learning
	\newblock \textit{TMM}, 2019.
	
	\bibitem[\protect\citeauthoryear{Chen}{2013}]{chen2013neil}
	X Chen et al,
	\newblock ``Neil: Extracting visual knowledge from web data.''
	\newblock \textit{ICCV}, 2013.	
	
	\bibitem[\protect\citeauthoryear{Chen}{2015}]{chen2015sense}
	X Chen et al,
	\newblock Sense discovery via co-clustering on images and text.
	\newblock \textit{CVPR}, 2015.
	
	\bibitem[\protect\citeauthoryear{Xu}{2016}]{xuicme2016}
	J Xu et al,
	\newblock Automatic Image Dataset Construction with Multiple Textual Metadata.
	\newblock \textit{ICME}, 2016.
	
	\bibitem[\protect\citeauthoryear{Shu}{2015}]{shu2015}
	X. Shu et al,
	\newblock ``Weakly-Shared Deep Transfer Networks for Heterogeneous-Domain Knowledge Propagation''	
	\newblock {\em ACM MM}, 2015.	

	\bibitem[\protect\citeauthoryear{Wan}{2009}]{wan2009latent}
	K Wan et al,
	\newblock A latent model for visual disambiguation of keyword-based image search.
	\newblock \textit{BMVC}, 2009.
	
	
	
	
	\bibitem[\protect\citeauthoryear{Min}{2016}]{min2016being}
	W Min et al,
	\newblock ``Being a Supercook: Joint Food Attributes and Multimodal Content Modeling for Recipe Retrieval and Exploration''
	\newblock {\em TMM}, 2017.
	
	\bibitem[\protect\citeauthoryear{Zhao}{2018}]{fang2018}
	F Zhao et al, 
	\newblock ``Dynamic Conditional Networks for Few-Shot Learning''
	\newblock \textit{ECCV}, 2018.
	
	\bibitem[\protect\citeauthoryear{Zhang}{2017}]{xie2017sde}
	G Xie et al, 
	\newblock ``SDE: A novel selective, discriminative and equalizing feature representation for visual recognition''
	\newblock {\em IJCV}, 2017.
	
	\bibitem[\protect\citeauthoryear{Xie}{2019}]{cvpr19aren}
	G Xie et al, 
	\newblock ``Attentive Region Embedding Network for Zero-shot Learning''
	\newblock {\em CVPR}, 2019.
	
	\bibitem[\protect\citeauthoryear{Shu}{2018}]{shu2018}
	X. Shu et al, 
	\newblock ``Hierarchical Long Short-Term Concurrent Memory for Human Interaction Recognition''
	\newblock {\em ArXiv:1811.00270}, 2018.
	
	\bibitem[\protect\citeauthoryear{Wang}{2015}]{wang2015instre}
	S Wang et al,
	\newblock ``INSTRE: A New Benchmark for Instance-Level Object Retrieval and Recognition''
	\newblock {\em TOMM}, 2015. 
	
	\bibitem[\protect\citeauthoryear{Hu}{2017}]{hu2017frankenstein}
	G Hu et al,
	\newblock ``Frankenstein: Learning deep face representations using small data''
	\newblock {\em TIP}, 2017.
	
	\bibitem[\protect\citeauthoryear{Hua}{2017}]{hu2017attribute}
	G Hu et al, 
	\newblock ``Attribute-enhanced face recognition with neural tensor fusion networks''
	\newblock {\em ICCV}, 2017.
	
	\bibitem[\protect\citeauthoryear{Liu}{2019}]{liutip2019}
	L Liu et al,
	\newblock Extracting Privileged Information for Enhancing Classifier Learning.
	\newblock \textit{TIP}, 2019.
	
	\bibitem[\protect\citeauthoryear{Tang}{2018}]{tangijcai2018}
	Z Tang et al,
	\newblock Extracting Privileged Information from Untagged Corpora for Classifier Learning.
	\newblock \textit{IJCAI}, 2018.
	
	\bibitem[\protect\citeauthoryear{Hua}{2017}]{huatmm2017}
	X Hua et al,
	\newblock Exploiting Web Images for Dataset Construction: A Domain Robust Approach.
	\newblock \textit{TMM}, 2017.	
	
	\bibitem[\protect\citeauthoryear{Miller}{1995}]{miller1995wordnet}
	G Miller. 
	\newblock Wordnet: a lexical database for english.
	\newblock \textit{Communications of the ACM}, 1995.
	
	\bibitem[\protect\citeauthoryear{Loeff}{2006}]{loeff2006discriminating}
	N Loeff et al,
	\newblock Discriminating image senses by clustering with multimodal features.
	\newblock \textit{ACL}, 2006.
	
	\bibitem[\protect\citeauthoryear{Saenko}{2009}]{saenko2009unsupervised}
	K Saenko et al,
	\newblock Unsupervised learning of visual sense models for polysemous words.
	\newblock \textit{NIPS}, 2009.
	
	\bibitem[\protect\citeauthoryear{Barnard}{2005}]{barnard2005word}
	K Barnard et al,
	\newblock ``Word sense disambiguation with pictures,''
	\newblock \textit{AI}, 2005.
	
	\bibitem[\protect\citeauthoryear{Lin}{2012}]{lin2012syntactic}
	Y Lin et al,
	\newblock Syntactic annotations for the google books ngram corpus.
	\newblock \textit{ACL}, 2012.
	
	\bibitem[\protect\citeauthoryear{Wang}{2014}]{wang2014tpami}
	X Wang et al,
	\newblock Web image re-ranking usingquery-specific semantic signatures.
	\newblock \textit{TPAMI}, 2014.
	
	\bibitem[\protect\citeauthoryear{Uijlings}{2013}]{selective2013}
	J.R. Uijlings et al,
	\newblock  Selective Search for Object Recognition
	\newblock \textit{IJCV}, 2013.
	
	\bibitem[\protect\citeauthoryear{Ren}{2015}]{rpn2015}
	S. Ren et al,
	\newblock Faster r-CNN: Towards real-time object detection with region proposal networks
	\newblock \textit{NIPS}, 2015.    
	
	\bibitem[\protect\citeauthoryear{Zhou}{2016}]{saliency2016}    
	B. Zhou et al,
	\newblock Learning deep features for discriminative localization
	\newblock \textit{CVPR}, 2016.

	\bibitem[\protect\citeauthoryear{Otsu}{1979}]{threshold1979}
	N. Otsu,
	\newblock A threshold selection method from gray-level histograms
	\newblock \textit{TCYB}, 1979.     
	
	\bibitem[\protect\citeauthoryear{Rumelhart}{1986}]{back1986}
	D Rumelhart et al,
	\newblock Learning representations by back-propagating errors.
	\newblock \textit{Nature}, 1986. 
	
	\bibitem[\protect\citeauthoryear{Divvala}{2014}]{divvala2014learning}
	S.~Divvala et al,
	\newblock ``Learning everything about anything: Webly-supervised visual concept
	learning,''
	\newblock {\em CVPR}, 2014. 
	
	\bibitem[\protect\citeauthoryear{Yang}{2019}]{prl2018}
	W Yang et al,
	\newblock ``Exploiting Textual and Visual Features for Image Categorization''	
	\newblock {\em PRL}, 2019.
	
	\bibitem[\protect\citeauthoryear{Liu}{2018}]{mta2018}
	H Liu et al,
	\newblock ``Deep Representation Learning for Road Detection using Siamese Network''	
	\newblock {\em MTA}, 2018.
	
	\bibitem[\protect\citeauthoryear{Huang}{2018}]{huangpu}
	P Huang et al,
	\newblock ``Collaborative Representation Based Local Discriminant Projection for Feature Extraction''	
	\newblock {\em DSP}, 2018.
	
	\bibitem[\protect\citeauthoryear{Zhang}{2016}]{zhang2016domain}
	J Zhang et al,
	\newblock A domain robust approach for image dataset construction.
	\newblock \textit{ACM MM}, 2016.
	
	\bibitem[\protect\citeauthoryear{Yao}{2019}]{yaotkde2019}
	Y Yao et al,
	\newblock Towards Automatic Construction of Diverse, High-quality Image Datasets.
	\newblock \textit{TKDE}, 2019. 
	
	\bibitem[\protect\citeauthoryear{Gella}{2016}]{acl2016}
	S. Gella et al,
	\newblock ``Unsupervised Visual Sense Disambiguation for Verbs using Multimodal Embeddings''
	\newblock {\em ACL}, 2016. 
	
	\bibitem[\protect\citeauthoryear{Lucchi}{2012}]{lucchi2012}
	A Lucchi et al,
	\newblock ``Joint image and word sense discrimination for image retrieval''
	\newblock {\em ECCV}, 2012.
	
	\bibitem[\protect\citeauthoryear{Golge}{2014}]{golge2014}
	E. Golge et al,
	\newblock ``Concept map: Mining noisy web data for concept learning''
	\newblock {\em ECCV}, 2014.
	
	\bibitem[\protect\citeauthoryear{Qiu}{2013}]{qiu2013}
	S. Qiu et al,
	\newblock ``Visual semantic complex network for web images''
	\newblock {\em ICCV}, 2013.
	
	\bibitem[\protect\citeauthoryear{Wah}{2011}]{branson}
	C. Wah et al,
	\newblock ``The Caltech-UCSD birds-200–2011 dataset''
	\newblock {\em Tech Report}, 2011.
	
	\bibitem[\protect\citeauthoryear{Lin}{2015}]{bilinear}
	T.-Y. Lin et al,
	\newblock ``Bilinear CNN models for fine-grained visual recognition''	
	\newblock {\em ICCV}, 2015.	
	
	\bibitem[\protect\citeauthoryear{Xu}{2017}]{xu2017}
	M Xu et al,
	\newblock ``Deep Learning for Person Reidentification Using Support Vector Machines''	
	\newblock {\em AIM}, 2017.
	
	\bibitem[\protect\citeauthoryear{Tang}{2017}]{tang2017}
	Z Tang et al,
	\newblock ``A New Web-supervised Method for Image Dataset Constructions''	
	\newblock {\em NEUROCOM}, 2017.
	
	\bibitem[\protect\citeauthoryear{Hua}{2016}]{mmm}
	X Hua et al,
	\newblock ``Extracting Visual Knowledge from the Internet: Making Sense of Image Data''	
	\newblock {\em MMM}, 2015.	
	
\end{thebibliography}
\end{document}